\documentclass[dvips]{article}
\usepackage{icrctc07}

\title{Energy Dependent Morphology in the PWN candidate HESS\,J1825--137}
\shorttitle{Energy Dependent Morphology in HESS\,J1825--137}

\authors{S. Funk$^{1}$, J.~A. Hinton$^{2}$,O.~C. deJager$^{3}$ for the
H.E.S.S.\ collaboration}
\shortauthors{Funk et al.}
\afiliations{
  $^1$Kavli Institute for Particle Astrophysics and
  Cosmology, SLAC, Menlo Park, CA-94025, USA\\ 
  $^2$ School of Physics and Astronomy, University of Leeds, Leeds LS2
  9JT, UK\\
  $^3$ Unit for Space Physics, North-West University, Potchefstroom
  2520, South Africa 
}

\email{Stefan.Funk@slac.stanford.edu}

\abstract{Observations with H.E.S.S.\ revealed a new source of very
high-energy (VHE) gamma-rays above 100 GeV -- HESS\,J1825--137 --
extending mainly to the south of the energetic pulsar
PSR\,B1823--13. A detailed spectral and morphological analysis of
HESS\,J1825--137 reveals for the first time in VHE gamma-ray astronomy
a steepening of the energy spectrum with increasing distance from the
pulsar. This behaviour can be understood by invoking radiative cooling
of the IC-Compton gamma-ray emitting electrons during their
propagation. In this scenario the vastly different sizes between the
VHE gamma-ray emitting region and the X-ray PWN associated with
PSR\,B1823--13 can be naturally explained by different cooling
timescales for the radiating electron populations. If this scenario is
correct, HESS\,J1825--137 can serve as a prototype for a whole class
of asymmetric PWN in which the X-rays are extended over a much smaller
angular scales than the gamma-rays and can help understanding recent
detections of X-ray PWN in systems such as HESS\,J1640--465 and
HESS\,J1813--178. The future GLAST satellite will probe lower electron
energies shedding further light on cooling and diffusion processes in
this source.}

\begin{document}
\maketitle
\section{Introduction}

The pulsar PSR\,B1823--13 and its surrounding X-ray pulsar wind nebula
(PWN) G18.0--0.2 is a system that has been studied by H.E.S.S.\ in
very high-energy gamma-rays above 200~GeV in unprecedented
detail~\cite{HESSJ1825II}. PWNe seem to constitute a significant
fraction of the population of identified Galactic VHE gamma-ray
sources detected by H.E.S.S.~\cite{FunkBarcelona} and as also
suggested by a statistical assessment of the correlation between
Galactic VHE $\gamma$-ray sources and energetic pulsars (see Carrigan
et al., these proceedings).  The gamma-ray emission in these objects
is typically thought to be generated by Inverse Compton scattering of
relativistic electrons accelerated in the termination shock of the
PWN.

Considering the population of VHE gamma-ray PWNe, HESS\,J1825--137 is
probably thus far the best example of the emerging class of so-called
\emph{offset Pulsar Wind nebulae} in which an extended VHE gamma-ray
emission surrounding an energetic pulsar is offset into one direction
of the pulsar. This offset is generally thought to arise from dense
molecular material in one direction of the pulsar that prevents an
symmetric expansion of the PWN (see e.g.~\cite{Blondin} for a
hydro-dynamical simulation and discussion of this effect).

As one of the best studied objects in VHE gamma-rays with an
observation time of nearly 70~hours, HESS\,J1825--137 has been used as
a template for the association of asymmetric PWN in VHE $\gamma$-rays
and X-rays~\cite{FunkBarcelona, YvesBarcelona}. In HESS\,J1825--137
the claimed association between the VHE $\gamma$-ray source and the
X-ray PWN rests on the following properties of the source:
\begin{itemize}
\item Same morphology (i.e.\ asymmetric extension to the south) in both
  bands but X-ray nebula much smaller ($\sim$ 5'') than $\gamma$-ray
  ($\sim$ 0.5$^{\circ}$) emission region
\item Spectral steepening of the VHE gamma-ray source away from the
  pulsar (i.e.\ decrease of gamma-ray extension with increasing
  energy). Interestingly the maximum of the VHE $\gamma$-ray emission
  is not coincident with the pulsar position but is shifted $\sim 17'$
  to the south-west.
\end{itemize}

The vastly different sizes of the emission region in the two wavebands
prevents at first glance a direct identification as a counterpart,
since the morphology can not be matched between X-rays and
gamma-rays. As will be explained in the following, the different sizes
can be explained in a time-dependent leptonic model by different
cooling timescales of the X-ray and of the VHE gamma-ray emitting
regions. Caution should however be used, if such an association serves
as a template for other unidentified H.E.S.S. VHE gamma-ray sources
with an energetic pulsar in the vicinity, in cases in which no X-ray
PWN has been detected so far.

\section{Observational data}

CO-Observations performed in the composite survey~\cite{Dame} show a
dense molecular cloud in the distance band between 3.5 and 4~kpc to
the north of PSR\,B1823--13 (located at $\sim 4$
kpc)~\cite{Lemiere}. This cloud seems to support the picture of an
offset PWN and could explain why the X-ray and VHE emission is shifted
to the south of the pulsar. Given the relatively high gamma-ray flux
and the rather large distance of the system of 4~kpc (in comparison to
the Crab), the required gamma-ray luminosity $L_\gamma \sim 3 \times
10^{35} \ \rm erg/s$ is comparable to the Crab luminosity. The
spin-down luminosity of the pulsar is, however, two orders of
magnitude lower than the Crab spin-down luminosity. Assuming the
distance of $\sim 4$~kpc is correct this shows that the efficiency of
converting spin-down power to gamma-ray luminosity must be much higher
than in the Crab Nebula, not unexpected, given the large magnetic
field in the Crab Nebula. Detailed time-dependent modelling of the
source shows indeed that (especially below $\sim 1$TeV) the energy
injection into the system must have been about an order of magnitude
higher in the past. Potentially the spin-down power of the pulsar was
significantly higher in the early stage of the pulsar evolution. For
the lower energy end of the H.E.S.S.\ spectrum and for modest magnetic
fields of a few $\mu$G as suggested by the large VHE gamma-ray flux,
the electron lifetimes become comparable to the pulsar age and
therefore ``relic'' electrons released in the early history of the
pulsar can survive until today and provide the required luminosity. It
should be noted that to this date no sensitive X-ray observation of
the region coinciding with the peak of the VHE gamma-ray emission has
been performed and a low surface-brightness extension to the south of
the X-ray PWN found by Gaensler et al.~\cite{Gaensler1825} remains an
interesting possiblility that should eventually be tested.

\begin{figure}
\begin{center}
\noindent
\fbox{\hbox{\vbox{\hsize=60mm \includegraphics
      [width=60mm]{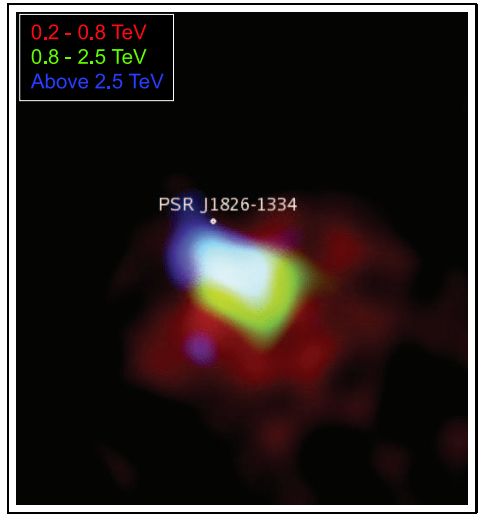} }}}
\end{center}
\caption{Three-colour image showing the gamma-ray emission in
  different energy bands (red: 0.2-0.8 TeV, green 0.8-2.5 TeV and
  blue: above 2.5 TeV). The different gamma-ray energy bands show a
  shrinking with increasing energy away from the pulsar
  PSR\,B1823--13.}\label{fig::HESS}
\end{figure}

\section{Energy dependent morphology}

\begin{figure*}[ht]
\begin{center}
\noindent 
\fbox{\hbox{\vbox{\hsize=150mm 
\includegraphics
    [width=150mm]{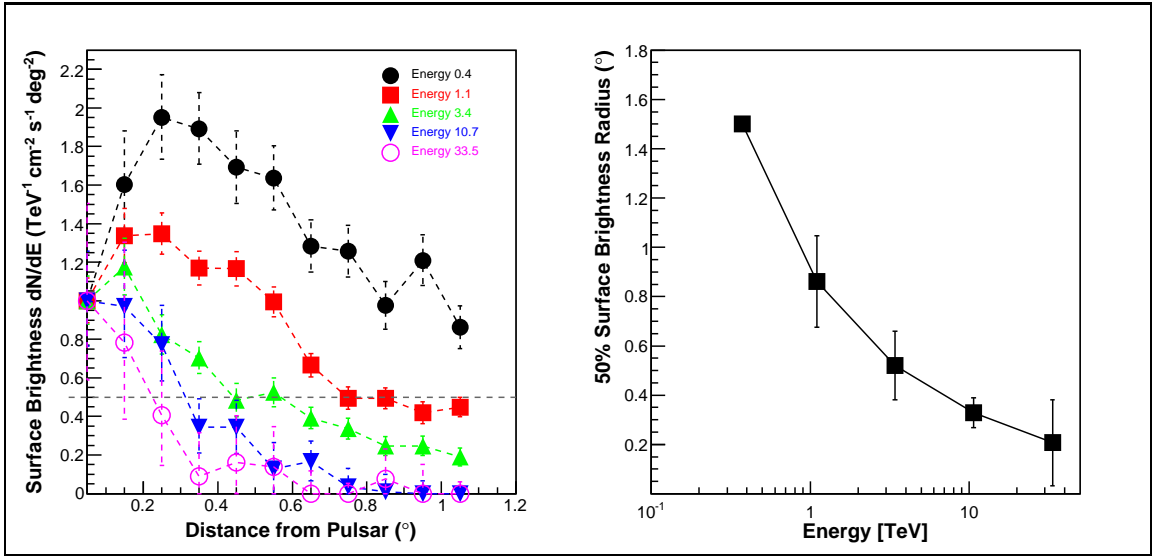}}}}
\end{center}
\caption{{\bf{Left:}} Surface brightness as a function of distance from
  the pulsar for different energy bands (derived from Figure 4 in
  Aharonian et al~\cite{HESSJ1825II}. The surface brightness is
  defined as the differential gamma-ray flux at a given energy scaled
  by the area of the extraction region and normalised by the flux for
  that energy at the pulsar position r = 0. {\bf{Right:}} Distance from
  the pulsar at which the surface brightness drops to 50\% of the flux
  at the pulsar position. The error bars are derived by fitting the
  falling points of the left plot, varying the fit parameters within
  the errors and recalculating the 50\% containment
  radius. }\label{fig::HESS2}
\end{figure*}

Given the large data set with nearly 20,000 $\gamma$-ray excess
events, a spatially resolved spectral analysis of HESS\,J1825--137
could be performed. For the the first time VHE $\gamma$-ray astronomy
an energy dependent morphology (see Figure~\ref{fig::HESS}) was
established~\cite{HESSJ1825II} in which the size of the emission
region decreases with increasing energy. This shrinking size with
increasing energy is equivalent to the statement of a steepening of
the spectral index away from the pulsar. The spectrum in
HESS\,J1825--137 changes from a rather hard photon index $\sim 2$
close to the pulsar to a softer value of $\sim 2.5$ at a distance of
$1^{\circ}$ away from the pulsar. Figure~\ref{fig::HESS2} shows the
surface brightness as a function of the distance from the pulsar for
different energy bands. Two clear trends are apparent in this figure:
a) the peak of the surface brightness shifts to lower energies (as
already suggested by the steepening of the energy spectrum away from
the pulsar) b) at low energies the surface brightness is nearly
independent of the distance whereas at the higher energies the surface
brightness drops rapidly with increasing distance from the pulsar. The
right panel of Figure~\ref{fig::HESS2} shows the derived radius
$R_{50}$ corresponding to the 50\% containment of the surface
brightness. This radius $R_{50}$ drops with increasing energy as
already apparent in Figure~\ref{fig::HESS}.

The steepening of the energy spectrum away from a central pulsar is a
property commonly observed in X-ray studies of PWNe other than the
Crab. For most of these system the total change in the photon index is
close to $\sim 0.5$ similar to what is seen in HESS\,J1825--137. It
should be noted that the results shown here represent the first
unambiguous detection of a spectral steepening at a fixed electron
energy (since the synchrotron emission seen in X-rays depends on the
magnetic field) in a PWN system. Spectral variation with distance from
the pulsar could result from (1) energy loss of particles during
propagation, with radiative cooling of electrons as the main loss
mechanism, from (2) energy dependent diffusion or convection speeds,
and from (3) variation of the shape of the injection spectrum with age
of the pulsar. Concerning (1): Loss mechanisms include amongst others
adiabatic expansion, ionisation loss, bremsstrahlung, synchrotron
losses and inverse Compton losses. Only synchrotron and IC losses can
result in a electron lifetime that decreases with increasing energy. A
source decreasing source size with increasing energy is therefore
generally seen as indicative of electrons as the radiating particles.

For continuous injection and short radiative lifetimes of the
electrons (in comparison to the age of the source), the spectral index
of the electrons changes by one unit as a result of the cooling,
yielding in a change of 0.5 in the the photon index. This matches
roughly what is seen in HESS\,J1825--137 when comparing the inner and
the outer nebula.  The lower energy gamma-rays (i.e. below $\sim 0.6$
TeV) correspond to mostly un-cooled low energy electrons (i.e. the
spectral index consistent with the injection spectral index). At these
low energies the electron lifetime becomes comparable to the age of
the source and the size is rather independent of the energy. At higher
energies the cooling break takes effect and the source size shrinks
with increasing energy as expected from electron cooling. The
XMM-Newton X-ray emitting electrons typically have much higher
energies ($\sim 100$~TeV) than the $\gamma$-ray emitting electrons
($\sim 10$~TeV), assuming a typical magnetic field of 5~$\mu$G. The
synchrotron cooling lifetime of X-ray emitting electrons is therefore
expected to be much smaller, resulting in a smaller spatial extension
in X-rays.

For systems like HESS\,J1825--137 a detailed study in X-rays trying to
detect the low surface brightness nebula in the soft X-ray band would
be very beneficial, is however very hard to achieve given the
absorption of soft X-rays.  The upcoming GLAST-satellite will observe
this object in a thus far rather unexplored energy regime especially
above $\sim 10$~GeV, where the angular resolution of the instrument
becomes comparable to the angular resolution of the ground-based
instrument. In this energy range GLAST will probe even lower energy
electrons and it will be interesting to compare the sizes of the GLAST
and the H.E.S.S.\ emission region. The H.E.S.S.\ results have shown
that a wealth of detail exists in gamma-rays at an angular scale of
$\sim 0.1^{\circ}$. Future instruments like CTA or AGIS might improve
this angular resolution even further.

\section{Acknowledgements}
"The support of the Namibian authorities and of the University of
Namibia in facilitating the construction and operation of H.E.S.S. is
gratefully acknowledged, as is the support by the German Ministry for
Education and Research (BMBF), the Max Planck Society, the French
Ministry for Research, the CNRS-IN2P3 and the Astroparticle
Interdisciplinary Programme of the CNRS, the U.K. Science and
Technology Facilities Council (STFC), the IPNP of the Charles
University, the Polish Ministry of Science and Higher Education, the
South African Department of Science and Technology and National
Research Foundation, and by the University of Namibia. We appreciate
the excellent work of the technical support staff in Berlin, Durham,
Hamburg, Heidelberg, Palaiseau, Paris, Saclay, and in Namibia in the
construction and operation of the equipment."

\bibliographystyle{plain}

\end{document}